\documentclass{PoS}

\title{Simulation of Scalar Field Theories with Complex Actions}

\ShortTitle{Simulation of Complex Actions}

\author{\speaker{Michael C. Ogilvie}\\
        Department of Physics, Washington University, St. Louis MO USA\\
        E-mail: \email{mco@wustl.edu}}

\author{Leandro Medina\\
        Department of Physics, Washington University, St. Louis MO USA\\
        E-mail: \email{leandro.medina.lv@gmail.com}}

\abstract{Many scalar field theory models with complex actions are invariant
under the antilinear ($PT$) symmetry operation $L^{\ast}(-\chi)=L(\chi)$.
Models in this class include the $i\phi^{3}$ model, the Bose gas
at finite density and Polyakov loop spin models at finite density.
This symmetry may be used to obtain a dual representation where weights
in the functional integral are real but not necessarily positive.
For a subclass of models satisfying a dual positive weight condition,
the partition function is manifestly positive.
The sign problem is eliminated; such models are easily simulated by a
simple local algorithm in any number of dimensions. Simulations of
models in this subclass show a rich set of behaviors. Propagators
may exhibit damped oscillations, indicating a clear violation of spectral
positivity. Pattern formation may also occur, with both
stripe and bubble morphologies possible.  
The existence of a positive representation is constrained
by Lee-Yang zeros: a positive representation cannot exist everywhere
in the neighborhood of such a zero.
Simulation results raise the possibility that pattern-forming behavior
may occur in finite density QCD in the vicinity of the critical line.
}

\FullConference{The 36th Annual International Symposium on Lattice Field Theory - LATTICE2018\\
		22-28 July, 2018\\
		Michigan State University, East Lansing, Michigan, USA.}

\begin{document}

\section{PT Symmetry and a New Algorithm for the Sign Problem}

The sign problem is known to be extremely difficult in the general
case. Within the general class of problems with complex weights, there
is a distinguished class of $PT$ symmetric models, which
includes many well-known models with sign problems, such as the $i\phi^{3}$
field theory and QCD at finite density. $PT$ symmetry refers
to any symmetry under the action of a discreet symmetry like parity
combined with complex conjugation. For a single scalar field $\chi$,
the action is invariant under
\begin{equation}
\mathcal{L}\left(\chi\right)=\mathcal{L}\left(-\chi\right)^{*}.
\end{equation}
This is referred to as a $PT$ symmetry because in the quantum-mechanical
systems in which the symmetry was initially studied, the field $\chi$
is naturally interpreted as a spatial coordinate $x$. In the case
of QCD or similar models at finite density, the symmetry is $CK$
where $C$ is charge conjugation and $K$ is complex
conjugation. It can be proven that such systems always have real representations
\cite{Meisinger:2012va}, and sometimes there are positive representations where all path integral
weights are real. In the latter case, the sign problem is completely
removed. 

The $PT$ symmetry condition can be used to find a natural
class of $PT$-symmetric scalar field models for which there
exists a representation where all weights are demonstrably positive.
The symmetry of $\mathcal{L}$ tells us that its Fourier tranforms
with respect to $\chi$ is a real function $\tilde{\mathcal{L}}\left(\tilde{\chi}\right)$;
this property extends to the action and to path integral weights.
Using Fourier transforms in the scalar field variables
can be understood as a kind of duality transform appropriate for real-valued
fields. We take the action to have the form
\begin{equation}
S(\chi)=\sum_{x}\left[\frac{1}{2}(\partial_{\mu}\chi(x))^{2}+V(\chi(x))-ih(x)\chi(x)\right]
\end{equation}
where $V\left(\chi\right)=V\left(-\chi\right)^{*}$. The external
field $h(x)$ is site-dependent, and used to deduce correlators
for $\chi$ in the new representation. 
We now rewrite the kinetic and potential terms as Fourier transforms.
The single-site kinetic term is easily written as a Fourier transform:
\begin{equation}
\exp\left[\frac{1}{2}\left(\partial\chi_{x}\right)^{2}\right]=\int d\pi_{\mu}(x)\exp\left[\frac{1}{2}\pi_{\mu}(x)^{2}+i\pi_{\mu}(x)\partial_{\mu}\chi_{x}\right]
\end{equation}
The weight term $w\left(\chi\left(x\right)\right)\equiv\exp\left[-V\left(\chi\left(x\right)\right)\right]$
associated with the potential term $V$ can be written as the transform
of a real dual weight $\tilde{w}\left(\tilde{\chi}\left(x\right)\right)$.
If $\tilde{w}>0$, then we can define the real dual potential $\tilde{V}\left(\tilde{\chi}\right)\equiv-\log[\tilde{w}\left(\tilde{\chi}\right)]$.
After a lattice integration by parts, it is straighforward to integrate
out $\chi$ to obtain the action
\begin{equation}
Z=\int\prod_{x}d\pi_{\mu}(x)\exp\left\{ -\sum_{x}\left[\frac{1}{2}\pi_{\mu}^{2}(x)+\tilde{V}(\partial\cdot\pi(x)-h(x))\right]\right\} .
\end{equation}
which is postive everywhere. In such a representation, standard lattice
simulation methods may be used. This representation is local and may
be used in any dimension; algorithmic implementation is easy. The
extension to multiple scalar fields is trivial.

\section{Models: positivity violation, pattern formation and the Bose gas}

We have found a rich set of models with two real scalar fields, $\chi$ and
$\phi$; the additional field $\phi$ transforms trivially under
$P$. We consider three different models. 
Our first two models have an imaginary
term $-ig\phi\chi$ as part of the potential, while the third, the
Bose gas at finite density, has an imaginary term in the kinetic part
of the action.

\begin{figure}
\includegraphics[height=3.5in]{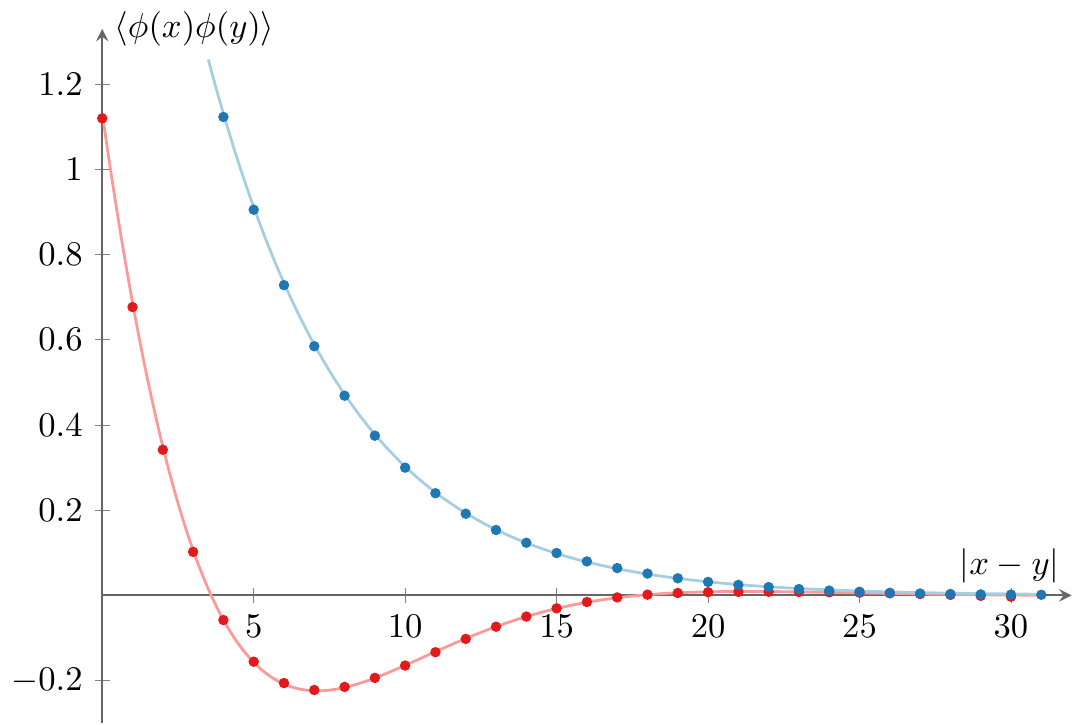}
\centering
\caption{\label{fig:icqprop}Propagator $\langle\phi(x)\phi(y)\rangle$ as
a function of $|x-y|$ for the $d=1$ ICQ model on a lattice with
$256$ sites. In both curves, $m_{\phi}^{2}=0.001$ and $g=0.1$.
The upper (blue) curve corresponds to $m_{\chi}^{2}=0.250$, while
the lower (red) curve has $m_{\chi}^{2}=0.002$. The lines represent
the analytical form of the continuum result; the errors bars on all
points are smaller than the points themselves.}
\end{figure}

Our first model is exactly solvable in any dimension but displays
interesting behavior as $g$ is varied \cite{Bender:2007wb,Bender:2016vdo}. 
This imaginary-coupled
quadratic (ICQ) model has a potential $V$ of the form $V(\phi,\chi)=m_{\phi}^{2}\phi^{2}/2+m_{\chi}^{2}\chi^{2}/2-ig\phi\chi$.
The eigenvalues of the mass matrix are given by either two real masses or a complex conjugate pair, as
required by the $PT$ symmetry of the model. The dual potential
takes the form $\tilde{V}(\phi,\partial\cdot\pi)=m_{\phi}^{2}\phi^{2}/2+(\partial\cdot\pi-g\phi)^{2}/2m_{\chi}^{2}$.
In figure \ref{fig:icqprop}, we show results for one-dimensional
simulations of the ICQ model in the two different regions.
The difference in behavior is striking between the
upper curve where there are two real masses, and the lower curve where
there is a complex conjugate mass pair. The lines represent the analytical
form of the continuum result for the propagators, and the points are from
simulations; the error bars
on the points are smaller than the points themselves. 

\begin{figure}
\includegraphics[height=4in]{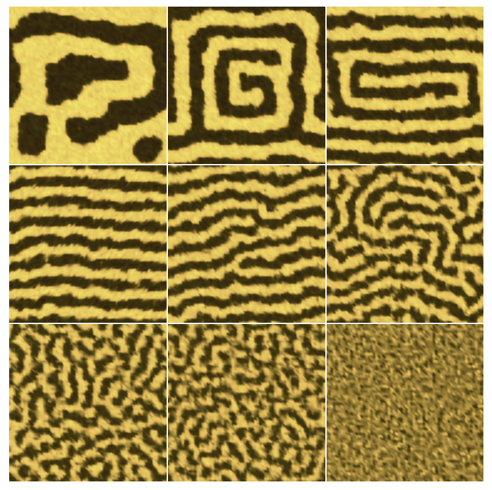}
\centering
\caption{\label{fig:Configuration-snapshots}Configuration snapshots of $\phi$
in the $d=2$ ICDW model on a $64^{2}$ lattice for several values
of $g$. From left to right, top to bottom: $g=0.8,\,0.9,\,1.0,\,1.1,\,1.2,\,1.3,\,1.4,\,1.5$
and $2.5$. The other parameters are $m_{\chi}^{2}=0.5$, $\lambda=0.1$
and $v=3$.}
\end{figure}

\begin{figure}
\includegraphics[height=4in]{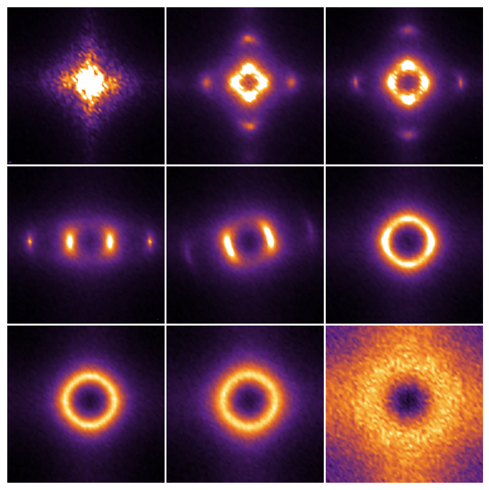}
\centering
\caption{\label{fig:Ring-of-fire}The Fourier transform $\tilde{\phi}\left(k\right)$
in the $d=2$ ICDW model on a $64^{2}$ lattice for several values
of$g$, averaged over 200 measurements. From left to right, top to
bottom: $g=0.8,\,0.9,\,1.0,\,1.1,\,1.2,\,1.3,\,1.4,\,1.5$ and $2.5$.
The other parameters are $m_{\chi}^{2}=0.5$, $\lambda=0.1$ and $v=3$.}
\end{figure}

Our second example generalizes the first, with a potential $V(\phi,\chi)=U(\phi)+m_{\chi}^{2}\chi^{2}/2-ig\chi\phi$
leading to $\tilde{V}(\phi,\partial\cdot\pi)=U(\phi)+(\partial\cdot\pi-g\phi)^{2}/2m_{\chi}^{2}$.
We consider here the case of the imaginary-coupled double
well (ICDW) model, where the potential has the form $U(\phi)=\lambda(\phi^{2}-v^{2})^{2}$.
Because the field $\chi$ enters quadratically, it may be integrated
out, yielding an effective action of the form 
\begin{equation}
S=\sum_{x}\left[\frac{1}{2}(\partial_{\mu}\phi(x))^{2}+U(\phi)\right]+\frac{g^{2}}{2}\sum_{x,y}\phi(x)\Delta(x-y)\phi(x)
\end{equation}
where $\Delta(x)$ is a free Euclidean propagator with mass $m_{\chi}$,
\emph{i.e.}, a Yukawa potential. This additional term in the action
acts to suppress spontaneous symmetry breaking. Models of this type
have been used to model a wide variety of physical systems and are
known to produce complicated pattern-forming phases 
\cite{Seul476,Nussinov:1999fu,Nussinov:2011wt}.
Constant solutions which extremize $S$ can be found by minimizing
$U(\phi)+g^{2}\phi^{2}/2m_{\chi}^{2}$. Linearizing the equation of
motion around such a solution $\phi_{0}$, the inverse propagator
is found to be 
\begin{equation}
p^{2}+U''\left(\phi_{0}\right)+\frac{g^{2}}{p^{2}+m_{\chi}^{2}}
\end{equation}
which is positive near $p^{2}=0$. However, if $U''\left(\phi_{0}\right)<0$,
the propagator can go negative at larger positive values of $p^{2}$,
indicating a tachyonic instability of the homogeneous state. In the
region of parameter space where this instability occurs, the equilibrium
phase is one where patterns form. This behavior is similar to that
found in related condensed matter physics models \cite{Seul476,Nussinov:1999fu,Nussinov:2011wt},
as well as in nuclear pasta models 
\cite{Ravenhall:1983uh,Hashimoto:1984xy,Horowitz:2008vf}.
Figure \ref{fig:Configuration-snapshots}
shows configuration snapshots on a $64^{2}$ lattice as $g$ is varied
within the tachyonic region. In figure \ref{fig:Ring-of-fire}, we
show the absolute value of the Fourier transform $\tilde{\phi}\left(k\right)$
for the same parameters. The ring
of enhanced modes, signaled by the appearance of tachyonic modes in
the homogenous solution propagator, is striking. 
This pattern-forming
behavior is independent of dimensionality; we have also observed it directly
in three-dimensional simulations. 

Our final two-component model, the Bose gas at finite density, is
different from the first three because the imaginary part of the action
is associated with the kinetic term in the action. The Bose gas at
finite density has the same anti-linear symmetry as QCD: the PT symmetry
is actually $CK$. The complex field $\Psi$ can be decomposed
as $\Psi=\frac{1}{\sqrt{2}}\left(\phi+i\chi\right).$
At zero chemical potential, this model is invariant under $\mathcal{C}$
and $\mathcal{K}$ separately but for chemical potential $\mu$ nonzero,
temporal derivative terms are complex and the Lagrangian is only $\mathcal{CK}$
symmetric: $\mathcal{L}\left(\phi,\chi\right)^{*}=\mathcal{L}\left(\phi,-\chi\right)$.
The action is given by
\begin{equation}
S(\chi)=\sum_{x}\left[\left(e^{\mu}\Psi^{*}\left(x+\hat{t}\right)-\Psi^{*}\left(x\right)\right)\left(e^{-\mu}\Psi\left(x+\hat{t}\right)-\Psi\left(x\right)\right)+\frac{1}{2}\left|\nabla\Psi(x)\right|^{2}+m^{2}\left|\Psi(x)\right|^{2}\right]
\end{equation}
After a series of Fourier transforms to eliminate $\chi$, the action $\tilde{S}$ takes the form
\begin{equation}
\tilde{S}=\frac{1}{2}\cosh\mu\,(\partial_{4}\phi)^{2}+\frac{1}{2}(\nabla\phi)^{2}+\frac{1}{2}\pi_{\mu}^{2}+\tilde{V}(\phi,\pi_{\mu})
\end{equation}
where 
\begin{equation}
\tilde{V}(\phi,\pi_{\mu})=\frac{1}{2}m^{2}\phi^{2}+\frac{1}{2M^{2}}\Bigl(\cosh\mu\,(\partial_{4}\pi_{4})+\nabla\cdot\vec{\pi}-\sinh\mu\left[\phi(x+\hat{e}_{4})-\phi(x-\hat{e}_{4})\right]\Bigr)^{2}
\end{equation}
with the mass parameter $M$ determined by $M^{2}\equiv2\left(1+\frac{1}{2}m^{2}-\cosh\mu\right)$.
The action is positive only if $M^{2}>0$. This is the lattice form
of a  familiar condition for scalar theories at finite density,
analogous to $m\le\mu$ in the continuum. The restriction to $M^{2}>0$
poses potential difficulties for an interacting theory with $m^2<0$. 
A dual representaton can be constructed
for $m^{2}<0$ using a polar decomposition of the fields \cite{Gattringer:2012df}.

\section{The case of $i\phi^{3}$ transitions}

Models in the $i\phi^{3}$ universality class are problematic because
their partition functions cannot be everywhere positive near an isolated
Lee-Yang zero. This directly implies that there can be no representation
of $Z$ which is everywhere positive near such a zero. The original
case of the Ising model in an imaginary field is instructive: the
partition function must vary in sign as Lee-Yang zeros are crossed.
According to the Lee-Yang circle theorem, such
a zero must lie on the line $Re(h)=0$, and $PT$ symmetry
tells us that $Z$ is real on this line. Because $Z$ is positive
when $Im(h)=0$, the partition function $Z$ must be negative along
$Re(h)=0$ between the first and second zero above the origin. 
This behavior has a natural explanation using $PT$ symmetry: zeros arise
when the largest eigenvalues of the transfer matrix $T_t$ form a conjugate
pair. If we suppose that the partition function can be written as
$Z=Tr\left(T_t^{N}\right)$ where $N$ is the extent of the system in
some direction, then $PT$ symmetry gives a representation
of $Z$ as 
\begin{equation}
Z=\sum_{p}\left(\lambda_{p}^{N}+\lambda_{p}^{*N}\right)+\sum_{r}\lambda_{r}^{N}
\end{equation}
where the sums are over all complex pairs $p$ of eigenvalues of $T$
and all real eigenvalues $r$. In the case where $N$ is large and
the partition function is dominated by a largest pair of complex eigenvalues,
$Z$ is approximately 
\begin{equation}
Z\approx\lambda_{0}^{N}+\lambda_{0}^{*N}=\left|\lambda_{0}\right|^{N}2\cos(N\phi_{0}).
\end{equation}
This indicates that $Z$ will oscillate between positive and negative
values as $N$ and $\phi_0$ vary whenever a complex pair of eigenvalues of $T$
dominates the partition function. 
A similar point of view on the relevance
of Lee-Yang zero has recently been given in \cite{deForcrand:2017rfp}.
There are several important $PT$-symmetric models which
contain within them forms of the $i\phi^{3}$ transition as the parameters
of the model are varied. Typical examples of such models are $Z(3)$
or $SU(3)$ spin systems which are strong-coupling approximations
to lattice QCD at nozero temperature and density. In $Z(3)$, for
example, the group elements are the set $\left\{ 1,\left(-1+i\sqrt{3}\right),\left(-1-i\sqrt{3}\right)\right\} $.
An additional term in the lattice action which favors the second
two elements can lead to a phase transition in the $i\phi^{3}$ universality
class.

\section{Conclusions}

There is a simple, local algorithm that allows for simulation of a
large class of scalar models with complex actions. This provides benchmark
results for proposed sign problem algorithms. Simulation indicates
that pattern formation may be a common feature of models with
sign problems and a phase transition, occurring when a tachyonic instability
of the global translation-invariant phase is present. The ICDW model
has several features in common with QCD at finite density. If we identify
$\phi$ with the real part of the Polyakov loop and $\chi$ with the
imaginary part, then the ICDW model as $g$ is varied is similar to motion along the critical
line of finite density QCD.
This suggests that there is a region of parameter
space where pattern formation occurs in models of finite density QCD.
Unfortunately, the presence of an $i\phi^{3}$ transition embedded
within a particular model\textquoteright s parameter space is an obstacle
to a successful transformation of many important sign problem models into a manifestly
positive form.


\end{document}